\documentclass[apj]{emulateapj}
\usepackage{natbib}
\usepackage{apjfonts}
\usepackage{adjustbox}
\usepackage{hyperref}

\def\d{{\rm d}}

\newcommand{\beq}{\begin{equation}}
\newcommand{\eeq}{\end{equation}}

\begin{document}

\title{Exploring the 2MASS {Extended} and {Point Source} Catalogues\\ with Clustering Redshifts} 

\author{
Mubdi Rahman\altaffilmark{1}, 
Brice M\'{e}nard\altaffilmark{1,2}
and Ryan Scranton\altaffilmark{3}
}
\altaffiltext{1}{Department of Physics and Astronomy, Johns Hopkins
  University, 3400 N. Charles Street, Baltimore, MD 21218}
\altaffiltext{2}{Kavli IPMU (WPI), the University of Tokyo, Kashiwa 277-8583, Japan}
\altaffiltext{3}{Department of Physics, University of California, One
  Shields Avenue, Davis, CA 95616, USA}
  \email{mubdi@pha.jhu.edu}

\begin{abstract}

The Two-Micron All-Sky Survey (2MASS) has mapped out the low-redshift Universe down to $K_S\sim14$ mag. As its near-infrared photometry primarily probes the featureless Rayleigh-Jeans tail of galaxy spectral energy distributions, colour-based redshift estimation is rather uninformative. Until now, redshift estimates for this dataset have relied on optical follow-up suffering from selection biases. Here we use the newly-developed technique of clustering-based redshift estimation to infer the redshift distribution of the 2MASS sources regardless of their optical properties. We characterise redshift distributions of objects from the \emph{Extended Source Catalogue} as a function of near-infrared colours and brightness and report some observed trends. We also apply the clustering redshift technique to dropout populations, sources with non-detections in one or more near-infrared bands, and present their redshift distributions. Combining all extended sources, we confirm with clustering redshifts that the distribution of this sample extends up to $z\sim0.35$. 
We perform a similar analysis with the \emph{Point Source Catalogue} and show that it can be separated into stellar and extragalactic contributions with galaxies reaching $z\sim0.7$. We estimate that the Point Source Catalogue contains 1.6 million extragalactic objects: as many as in the Extended Source Catalogue but probing a cosmic volume ten times larger.
\end{abstract}


\section{Introduction}
\label{sect:intro}

\begin{figure*}[!ht]
\begin{center}
\includegraphics[scale=0.8]{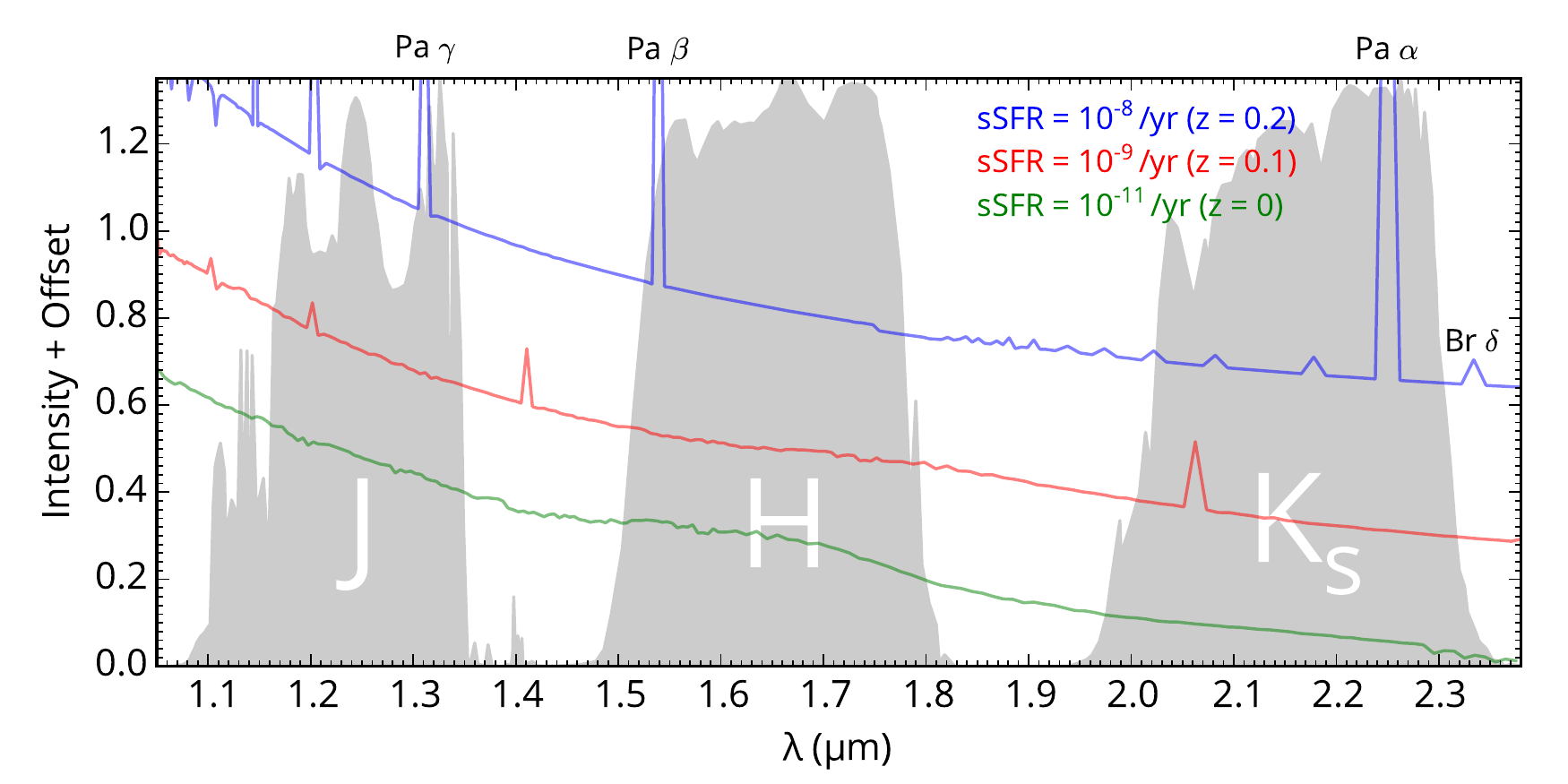}
\caption{
    Example spectral energy distributions of galaxies in the NIR. Model galaxy SEDs are produced by the GALEV simulation suite \citep{kotulla09}. Models with different specific star formation rates (sSFR) are shown with arbitrary offset and redshifted as indicated, with prominent emission lines labelled. The green curve is representative of elliptical galaxies with older stellar populations, while the blue curve is representative of a starburst. For reference, the 2MASS filter response curves are shown in the background. There are only a limited number of emission lines available for any spectroscopic redshift determination. Further, the SEDs of the galaxy are smooth, limiting the amount of information available for photometric redshifts.}
    \label{fig:sed}
\end{center}
\end{figure*}

The Near-infrared (NIR) sky provides a view of the Universe less affected by dust obscuration than in the visible wavelengths. Observations in this regime significantly reduce extinction effects due to the Milky Way as well as the self-obscuration of extragalactic sources, particularly when viewed edge-on. The Two-Micron All-Sky Survey \citep[2MASS;][]{skrutskie06} has produced the largest compilation of sources detected in the near-infrared ($1-2.4\,\mu$m) containing over 470 million sources in the Point Source Catalogue (PSC; primarily stars) and 1.6 million sources in the Extended Source Catalogue (XSC; primarily galaxies, \citealt{jarrett00}). The full-sky coverage makes this survey particularly well suited for galaxy studies and cosmological tests limited by cosmic variance. However, such extragalactic experiments typically require the knowledge of galaxy redshifts; unfortunately, photometric redshift estimation based only on near-infrared data is difficult. At low redshift, most flux received from galaxies originates from the Rayleigh-Jeans tail of stellar photospheres, a featureless spectral energy distribution. This limits the information available to discriminate redshift, age or metallicity. 

Photometric redshift estimates based solely on near-infrared data have been attempted \citep{jarrett04,kochanek03}. Since no colour variation is expected from Rayleigh-Jeans tails, redshift estimation is typically performed by assuming a fixed K-band galaxy luminosity and directly inferring distance based on observed flux. Such crude estimates typically lead to redshift errors of order $50\%$. Recently, several authors have cross-matched 2MASS galaxies with optical counterparts: \citet{way06}, \citet{way09}, \citet{wang08}, and \citet{wang09}, using the Sloan Digital Sky Survey \citep[SDSS;][]{york00} and \citet{francis2010} using the SuperCOSMOS survey \citep{hambly01} to estimate redshifts. \citet{bilicki14} also use SuperCOSMOS as well as WISE observations in the infrared to better constrain the redshift. Generally, they show that for typical 2MASS sources, when optical photometry is available, the near-infrared data does not substantially improve the accuracy of redshift estimation. While optical data brings in useful information on redshift estimation, this approach suffers from one limitation: a selection bias is introduced by the requirement of optical data as certain near-infrared sources will not be detected in the optical; such sources are referred to as optical ``dropouts''. 

Direct redshift estimation based on near-infrared spectroscopic observations also faces important limitations, arising from the discontinuous transparency of the Earth's atmosphere together with the small number of strong spectral features located in the near-infrared (see Figure \ref{fig:sed}). Similar to the photometric approach, in practice the characterization of near-infrared source redshifts has relied on additional spectroscopic data from optical wavelengths. The 2MASS Redshift Survey \citep[2MRS,][]{huchra12} has produced optically-based spectroscopic redshifts for about $45\,000$ flux-limited sources. This corresponds to only 3$\%$ of its parent extended source catalogue and is restricted to the nearby Universe ($z < 0.1$). Similarly, the 6dFGS redshift survey provides a deeper completeness of sources over the southern sky \citep{jones09}. Another, independent, source of spectroscopic redshift information comes from cross-matching 2MASS sources to the Sloan Digital Sky Survey. At $K_s<14$, about $90\%$ of the 2MASS extended sources overlapping the SDSS footprint have optical spectra and redshift measurements \citep{strauss02}. The remaining 10\% correspond to either sources lost in the targeting of the SDSS spectra and to optical ``dropouts'': sources not detected by the SDSS photometry, either from extreme reddening or unusual spectral energy distributions. Consequently, the SDSS spectroscopic redshift estimates of near-infrared sources also suffer from optical selection biases.

In order to circumvent this limitation affecting both photometric and spectroscopic samples, one needs to estimate the redshift distribution of near-infrared sources independently of optical detectability. This can be conducted using clustering-based redshift estimation. Rather than using spectral energy distributions, this technique infers redshifts from angular clustering measurements. This approach does not require any additional photometric information on the selected sources. It can therefore be applied to the entire 2MASS survey irrespective of optical counterparts. The use of spatial clustering to extract redshift information goes back to \citet{seldner79}. While the idea had been known for decades, theoretical and practical approaches to clustering-based redshift estimation followed much later \citep{landy96, ho08, newman08, menard13,schmidt13,mcquinn13}. The feasibility and accuracy of this clustering-based redshift inference with data from the SDSS has been investigated by our team \citep[][]{rahman15}. Here we use this technique to explore and characterise the extragalactic sources of 2MASS, in both the extended and point source catalogue down to the limiting magnitude of the survey.

\section{Data Analysis}
\label{sect:dataanalysis}

\subsection{Clustering-based Redshift Estimation}
\label{subsect:clustz}

Our approach is based on the method introduced in \citet{menard13}, tested against simulations in \citet{schmidt13}, and fully implemented and applied to data in \citet{rahman15}. We refer the reader to these papers for the detailed description of the formalism and technical considerations. In this section, we briefly re-introduce the main concepts.

We consider two populations of extragalactic objects: (i) a \emph{reference} population for which the angular positions and redshifts of each object are known. This population is characterised by a redshift distribution $\d{\rm N_r}/\d z$, a mean surface density $n_r$, a total number of sources $N_r$, and a clustering amplitude or \emph{bias} $b_r$; and (ii) an \emph{unknown} population for which angular positions are known but redshifts are not. Similarly, this population is characterised by the quantities $\d{\rm N_u}/\d z$, $n_u$, $N_u$ and $b_u$. The basic principle is that if the two populations do not overlap in redshift, their angular correlation is expected to be zero (ignoring gravitational lensing effects). As discussed by \citet{menard13}, in the ideal case of an unknown sample located within a narrow redshift range, one can probe its redshift distribution by splitting the reference population into contiguous redshift slices $\delta z_i$ and measuring the angular or spatial correlations with the unknown population $w_{ur}(\theta,z_i)$ for each subsample $i$. Once a cross-correlation signal is found, the amplitude of the redshift distribution is simply obtained through the normalization:
\begin{equation}
\int \d z\;{\rm d N_u}/{\d z} = {\rm N_u}. 
\label{eq:normalization}
\end{equation}
This normalization alleviates the need to characterise the amplitude of the clustering bias $b_u$.

The redshifts derived through this method use information solely from a source's angular position rather than the source's spectral energy distribution (SED). Consequently, it can be used for objects detected in only one bandpass. 
Classical photometric redshift estimation using SED fitting suffers from well-documented limitations, such as a need for a complete set of spectral templates, catastrophic failures, and dust reddening effects. Machine learning-based photometric redshift estimation has also been used extensively, but has similar limitations of catastrophic failures and dust reddening, and can be significantly affected by incomplete or biased training sets. In contrast, the fundamental limitation of clustering redshifts is the degeneracy between the redshift distribution of the unknown population $\d{\rm N_u}/\d z$ and the redshift evolution of its bias $b_u(z)$. As the width of the redshift distribution broadens, this introduces errors in the estimation, but as motivated theoretically in \citet{menard13} and demonstrated with real data by \citet{rahman15}, such errors can be made sufficiently small for a large range of astrophysical applications. We note that only the evolution of the bias with redshift imprints an effect on the clustering redshift, and not its specific value. In addition, this effect is minimised by subdividing the unknown sources into subsamples that, by construction, will have redshift distributions narrower than that of the total sample. This technique has been verified by testing against spectroscopic galaxies \citep{rahman15}, and has been used to measure the redshift distribution of the entire Sloan Digital Sky Survey photometric catalogue \citep{rahman15b}. The SDSS work demonstrates that clustering and photometric redshifts have unrelated systematics, and in cases where photometric redshifts are unsuitable, clustering redshifts can provide an alternative path to inferring redshift information. 

We estimate clustering redshift distributions with the procedure implemented in \citet{rahman15}. We refer the reader to this paper for the details of the implementation. Here we simply list the key parameters used in the present analysis: our reference spectroscopic sample is constructed by combining the SDSS Legacy spectroscopic sample extending up to $z \simeq 0.45$ \citep{strauss02,eisenstein01}, and the CMASS luminous red galaxies from the Baryon Oscillation Spectroscopic Survey extending to $z \simeq 0.8$ \citep{padman12}. This sample has been used to maximize the available number of sources for cross-correlation, consequently minimizing measurement noise. For this reference sample, we use the integrated bias evolution presented in \citet{rahman15} and use $d\log \overline{b}_u/d\log z = 1$ for the unknown population. The error induced by this choice of bias is discussed at length in \citet{menard13}. We measure the angular overdensity in an aperture of $0.3 < r < 3$~Mpc weighted by $\theta^{-0.8}$. To avoid measuring correlation functions over very large angular apertures on the sky and limit cosmic variance sampling limitations, we restrict our analysis to redshifts greater than $z=0.03$. This limit is chosen due to the systematics (i.e., excluded areas, varying background densities) introduced in the measurement of source densities over large areas on sky.

\subsection{The 2MASS Survey}

Throughout this work, we use the 2MASS Survey and the 2MASS Redshift Survey, which we describe here. Near-infrared sources detected in the 2MASS survey are divided into two samples: the Extended Source Catalogue (XSC) containing mostly galaxies, and the Point Source Catalogue (PSC) containing mostly stars. Based on the extended sources, an additional useful value-added catalogue is available: the 2MASS Redshift Survey that contains spectroscopic redshifts for a flux-limited sample. 

\begin{figure}[t]
\includegraphics[scale=0.7]{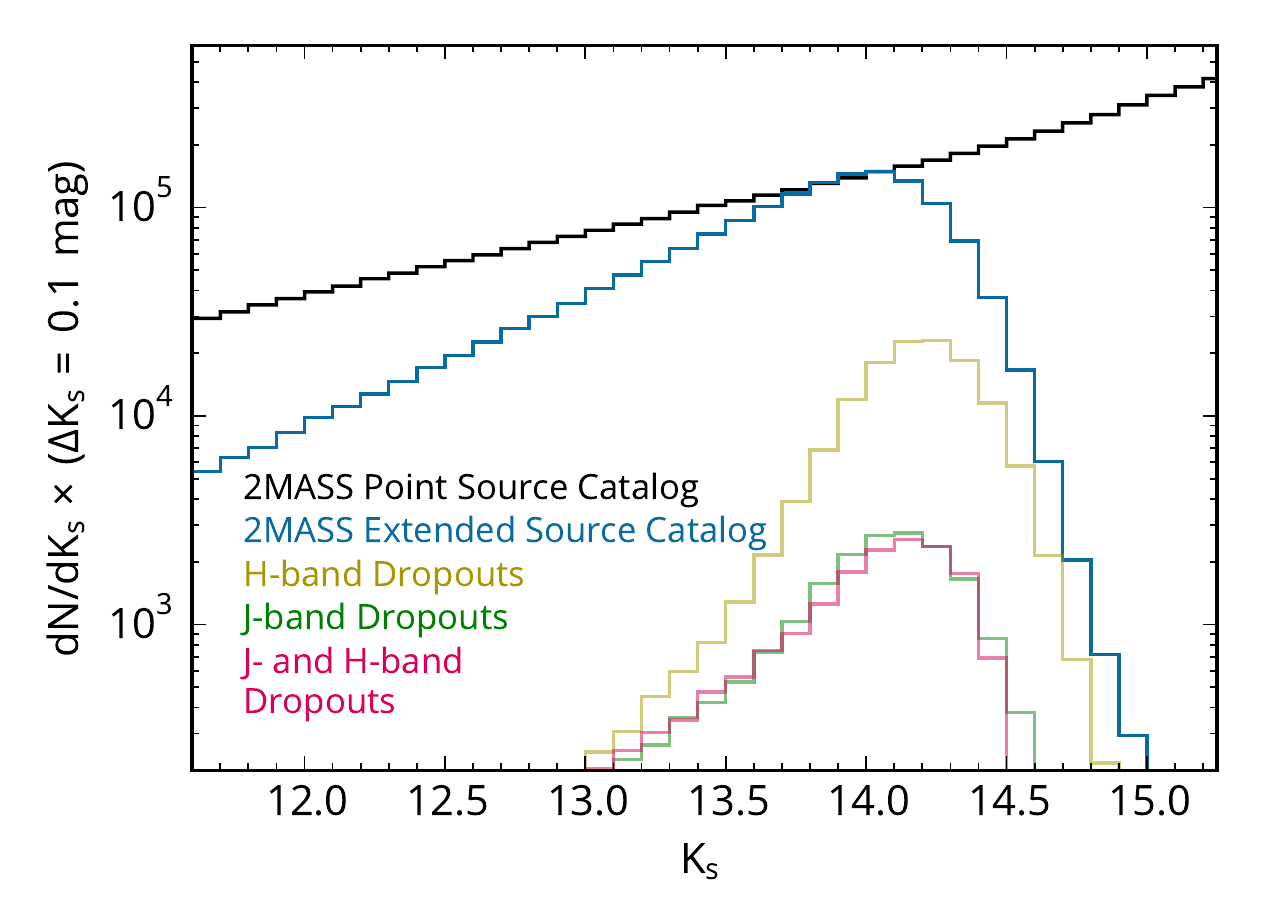}
\caption{
	The magnitude distribution of the 2MASS Point Source Catalogue in the Northern Galactic cap (black) and Extended Source Catalogue (blue). For the extended sources, the contribution of sources without J-band (green) or H-band detections (yellow), and both H- and J-band magnitudes (magenta) are also indicated. The similarity in density of point ane extended sources at K$_{\textrm s}$ = 14.0 is purely coincidental.
	\label{fig:magdist}
}
\end{figure}

The extended source catalogue contains all sources with an angular size larger than the 2MASS PSF (3\arcsec{} in the K$_{\textrm s}$ band) with a minimum signal-to-noise ratio of 7 in any of the three 2MASS bands \citep{skrutskie06}. In practice, however, this results in a required detection in the K-band for the majority of the objects. The catalogue consists of extragalactic sources that dominate at high Galactic latitude, together with a fraction of Galactic sources closer to the midplane.
Extended sources consistent with double or multiple stellar system are excluded through colour information\footnote{Through the use of the ``g-score'' as described in the 2MASS Explanatory Supplement: \url{http://www.ipac.caltech.edu/2mass/releases/allsky/doc/sec2_3b.html}}. At its photometric limits, the extended sources contains objects without measured magnitudes in the J and/or H bands, which we refer to as ``dropouts''. The completeness of the extended source catalogue is estimated to be $>95\%$ at K$_{\textrm s} < 14.0$, as determined from a comparison analysis to the Virgo Cluster\footnote{Explanatory Supplement to the 2MASS All Sky Data Release and Extended Mission Products: \url{http://www.ipac.caltech.edu/2mass/releases/allsky/doc/sec6_5b1.html}}. The 2MASS (Vega-based) magnitudes we use for extended sources in this paper are the fiducial photometry with radii set by the 20 mag/arcsec$^2$ isophot in the K$_{\textrm s}$-band. We present the magnitude distribution of the catalogue in Figure \ref{fig:magdist}. 

The remainder of 2MASS detected sources fall into the Point Source Catalogue. The vast majority (90\%) of the catalogue is located within 30\degr{} of the Galactic midplane and is dominated by stars. The sources at higher galactic latitude are expected to consist of both stars and galaxies with angular diameters smaller than the 2MASS point source function. We present the magnitude distribution of a Northern Galactic cap sample of the PSC in Figure \ref{fig:magdist}. The PSC is complete down to K$_{\textrm s} < 14.3$\footnote{Explanatory Supplement to the 2MASS All Sky Data Release and Extended Mission Products: \url{http://www.ipac.caltech.edu/2mass/releases/allsky/doc/sec6_5a1.html}}.

A subsample of 2MASS galaxies has been selected for spectroscopic redshift follow-up through the 2MASS Redshift Survey \citep[2MRS;][]{huchra12}. This sample is 98\% complete to K$_{\textrm s}$ = 11.75 over 91\% of the sky. Optical spectroscopy of these sources was taken at a combination of observatories in both the Northern and Southern hemisphere over 14 years to ensure near-all sky coverage. Augmenting their observed redshifts with those from complementary surveys, the 2MRS Catalogue consists of 44 599 spectroscopically measured redshifts. While this work marks a substantive increase in the redshift information available for infrared-selected galaxies, this amounts to only 3\% of the entire 2MASS extended source catalogue and does not provide any information for the minimal number of sources not detected in the optical at this flux limit.

To ensure homogeneous coverage of the reference sample and to minimise the effect of both Galactic extinction and sources in the 2MASS extended source catalogue, our analysis focuses on a 4800 square degree area within the Northern Galactic Cap, defined by:
\begin{eqnarray}
131\degr < &\alpha& < 241\degr\nonumber\\
5\degr < &\delta& <  60\degr \label{eq:footprint}
\end{eqnarray}
with fields surrounding bright stars removed. We use no other criteria or flags for the sample to avoid placing additional (possibly complicated) selection functions on the sample. Additionally, we do not make any adjustments for source extinction from external information. To account for the fluctuation of the source density due to cosmic variance and other potential systematic effects, we estimate the mean source density more locally by measuring it independently into 16 equal area regions spanning the entire footprint.

\subsection{Testing clustering redshifts\\ with 2MASS spectroscopic sources}
\label{subsect:2mrs}

\begin{figure}[t]
\center
\includegraphics[scale=0.7]{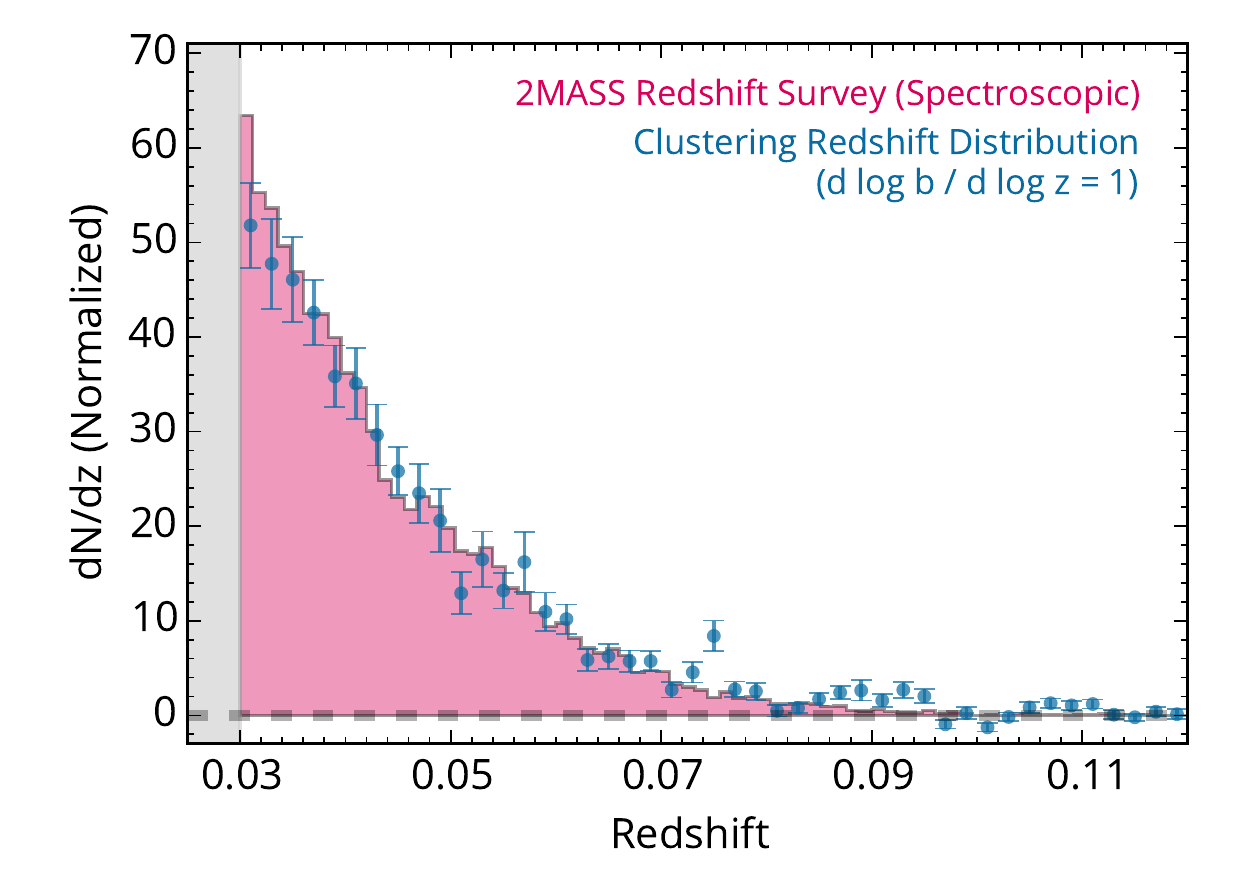}
\caption{
	A comparison between the spectroscopic redshifts (pink) and clustering redshifts (blue) of galaxies selected in the 2MASS Redshift Survey \citep{huchra12}. For the clustering redshifts, we use $d\log \overline{b}/d \log z = 1$, which is used in \citet{rahman15}.  
	\label{fig:2mrscrd} }
\end{figure}

\begin{figure*}[t]
\includegraphics[scale=1.0]{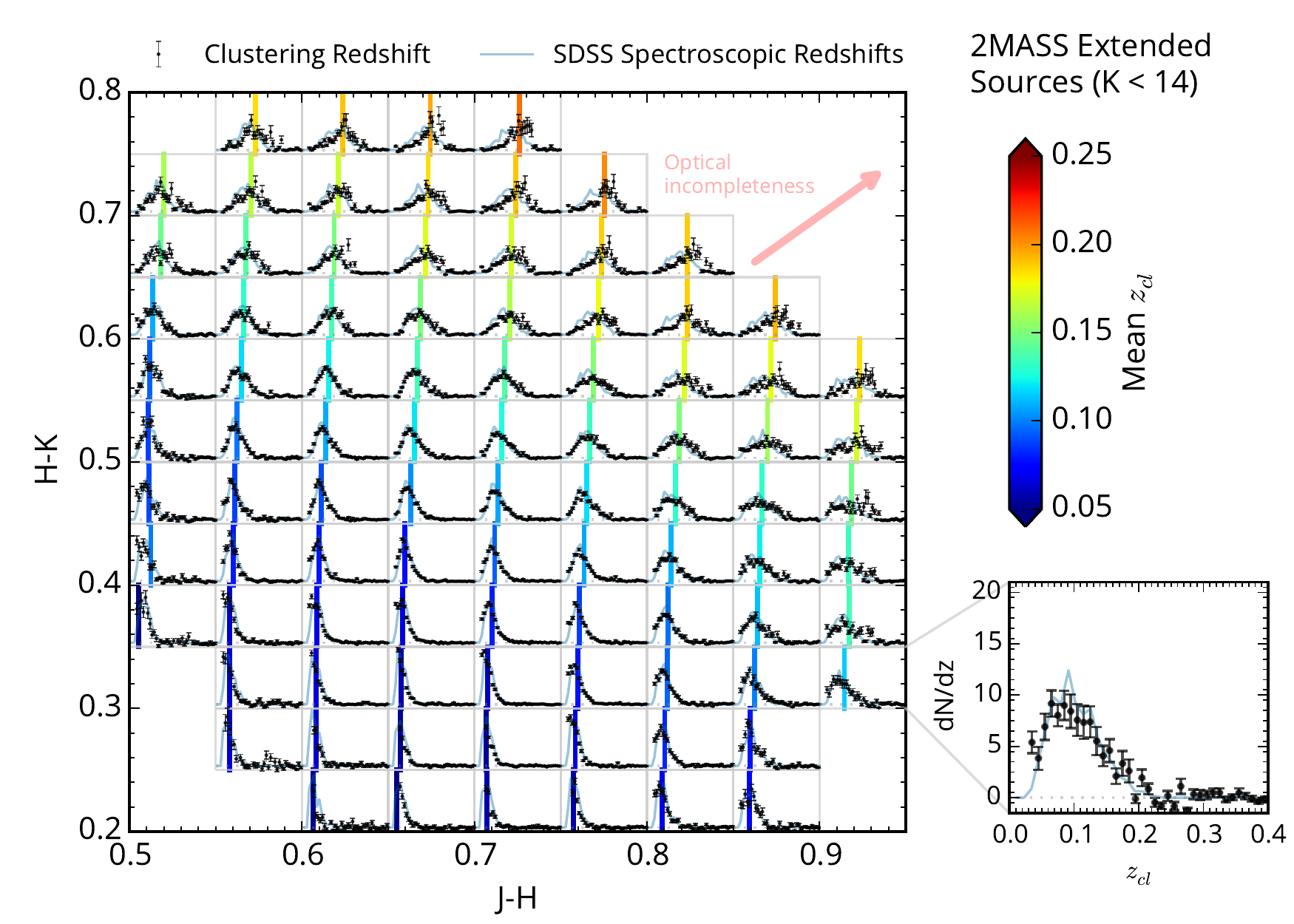}
\caption{
	The full colour-separated redshift distribution of the 2MASS Extended Source Catalogue with K$_{\textrm{s}} < 14$. Each cell represents the redshift distribution of the sources within the designated J-H and H-K$_{\textrm s}$ range. Each redshift range spans $0.03 < z < 0.4$, and is normalised by the number of sources within the range. The mean of each distribution is indicated by the coloured vertical line, with redshift values indicated in the colour bar.  The overall redshift of the distributions increases with colour, as fiducially expected. The redshift distribution of the 2MASS galaxies with spectra available from SDSS is indicated in the background of each cell. The red arrow indicates the direction in colour space where the SDSS spectroscopic sample will be biased towards brighter and more nearby objects with respect to the full 2MASS sample. 
	\label{fig:colspace}
}
\end{figure*}

We first test the robustness of our clustering redshift technique in the context of 2MASS data. To do so we estimate clustering redshifts for sources in the 2MASS Redshift Survey (with flux limit $K_s < 11.75$) where complete spectroscopic redshift measurements are available, over the sky footprint presented in Eq. \ref{eq:footprint}. This particular set of galaxies can be treated as a validation of the technique on a flux-limited subsample of the data. The redshift range spanned by these galaxies is sufficiently narrow ($\Delta z\sim 0.1$) that the redshift evolution of the galaxy bias can be neglected. We therefore apply the clustering redshift technique to the sample as a whole, without subsampling in photometric space. Figure~\ref{fig:2mrscrd} shows the distribution of spectroscopic redshifts for this sample with the purple histogram and the distribution of clustering redshifts with the black data points, binned with $\Delta z = 0.002$. 
The two distributions show good agreement, verifying the robustness of our clustering redshift estimation method. This test also shows that clustering redshift estimation is feasible with sparse samples: the source density of the 2MASS spectroscopic sample is about one object per square degree.

\section{Clustering Redshifts of 2MASS\\ {extended} sources}
\label{sect:results}

We now estimate clustering redshift distributions for the entire 2MASS XSC with $K_S<14$. Since this sample is expected to span a redshift range substantially larger than that of the 2MASS redshift survey, the degeneracy between its galaxy bias and redshift distribution may have an appreciable effect on the accuracy of the redshift distribution estimate (as discussed in \S\ref{subsect:clustz}). 
To minimise it, we first subsample the dataset in colour space to reduce the different galaxy types within each individual subsample, thus reducing the degeneracy between the galaxy bias and redshift distribution. We characterise the redshift distribution of objects selected in each colour sample, including near-infrared dropouts. The redshift distributions of the individual colour-based subsamples will be combined to present the global clustering redshift distribution of the 2MASS extended sources.

\subsection{Sampling the near-infrared colour space}
\label{sect:colourspace}

We select sources in square $(J-H,H-K_S)$ colour cells with a width of 0.05 mag. The sise of the cells is chosen to ensure the maximum number of colour cells while maintaining sufficient on-sky densities to measure the angular cross-correlation. We note that this binning is narrower than the typical colour error of most 2MASS sources.  We restrict our analysis to cells with source densities greater than 0.06 per square degree, which we have found to be the minimum density required for accurate measurement of mean density; this corresponds to about 90\% of the 2MASS extended source catalogue. The corresponding cells are displayed in Figure~\ref{fig:colspace}. In each cell we measure the distribution of clustering redshifts over the range provided by our reference sample, ($0.03 < z < 0.8$). Since we do not detect any signal at $z \geq 0.35$, we show clustering redshift measurements only up to $z=0.4$. Each redshift distribution is normalised to unity (i.e., $\int {dN}/{dz}\;dz = 1$). Errors are estimated through Poisson statistics. 

Spanning the colour space, we can observe redshift distributions from $z\sim0$ to about $0.4$, with mean redshifts ranging from $0.05$ to $0.25$. For bluer sources we observe narrower redshift distributions with a mean redshift of $z\sim0.1$ and a width of about $\sigma_z=0.05$. To visually display relationships between near-infrared colours and redshift, we show the mean redshift of each colour cell using a coloured vertical bar. As can be seen, the relationship between $H-K$ colour and redshift is steeper than that seen for the $J-H$ colour. The $J-H$ colour is more degenerate with the stellar composition of a galaxy than the $H-K$ colour; the H- and K-bands being further down the Rayleigh-Jeans tail in the rest frame, flux from galaxies at higher redshift will still arise from the Rayleigh-Jeans tail in these bands. Further, as the colours become redder, the redshift distributions become wider as well. This phenomenon likely comes from the colour-redshift degeneracy between higher redshift, dust-poor galaxies and lower redshift dusty galaxies. The results presented in this figure illustrate the power of clustering redshift estimation; this technique allows us to characterise redshift distributions in a photometric space (near-infrared colours) where classical photometric redshift techniques fail from the lack of strong correlation between colour and redshift \citep{jarrett04}. Our analysis also demonstrates that \emph{some} redshift information can be extracted from the near-infrared colours of the 2MASS galaxies.

In order to verify the validity of our clustering redshift estimates, we can compare our results to (incomplete) spectroscopic observations in the optical.
The SDSS legacy spectroscopy has observed about 90\% of the 2MASS galaxies that lie within its footprint. In the same figure we present the spectroscopic redshift distributions of sources selected as a function of near-infrared colours, using thin blue lines. As can be seen, for the vast majority of the colour space, clustering and spectroscopic redshift estimates are in good agreement. This further validates our method and illustrates the level of accuracy reachable by clustering redshifts ($>95\%$ over the full colour space). It is interesting to point out the existence of small differences between the spectroscopic and clustering redshifts, gradually rising toward the upper-right corner of the figure. These discrepancies appear to be caused by incompleteness of the spectroscopic sample. As the NIR colour increases, the spectroscopic sample is limited to  brighter and therefore lower redshift sources, whereas the clustering redshift distribution uses all sources. The upper-right corner of the figure (high $J-H$ and high $H-K$) corresponds to the reddest near-infrared sources, which are more likely missing optical detections. The difference between the spectroscopic and clustering redshift estimates mostly reflects the selection bias introduced by the requirement of optical data, as discussed in the Introduction. This illustrates the bias implicit in using optical data to characterise the redshift distribution of NIR sources and the advantage of clustering redshift estimation.

Since this analysis maps the magnitude and colour of a 2MASS galaxy onto a redshift distribution, this information can be used as a NIR-only photometric redshift, entirely unbiased by any models or training sets. We make the data and code for the full redshift distributions available online\footnote{The data and code to access the full redshift distributions is available at \url{http://www.pha.jhu.edu/~mubdi/2massz}}. 
We note that the redshift distribution corresponding to a given cell in colour space can be interpreted in two ways: it provides us with an estimate of the redshift distribution of the population of objects living in that cell or, alternatively, the probability distribution function for the redshift of an individual galaxy from the 2MASS survey living in that cell.

\subsection{Near-infrared dropouts}

\begin{figure}
\includegraphics[scale=1.0]{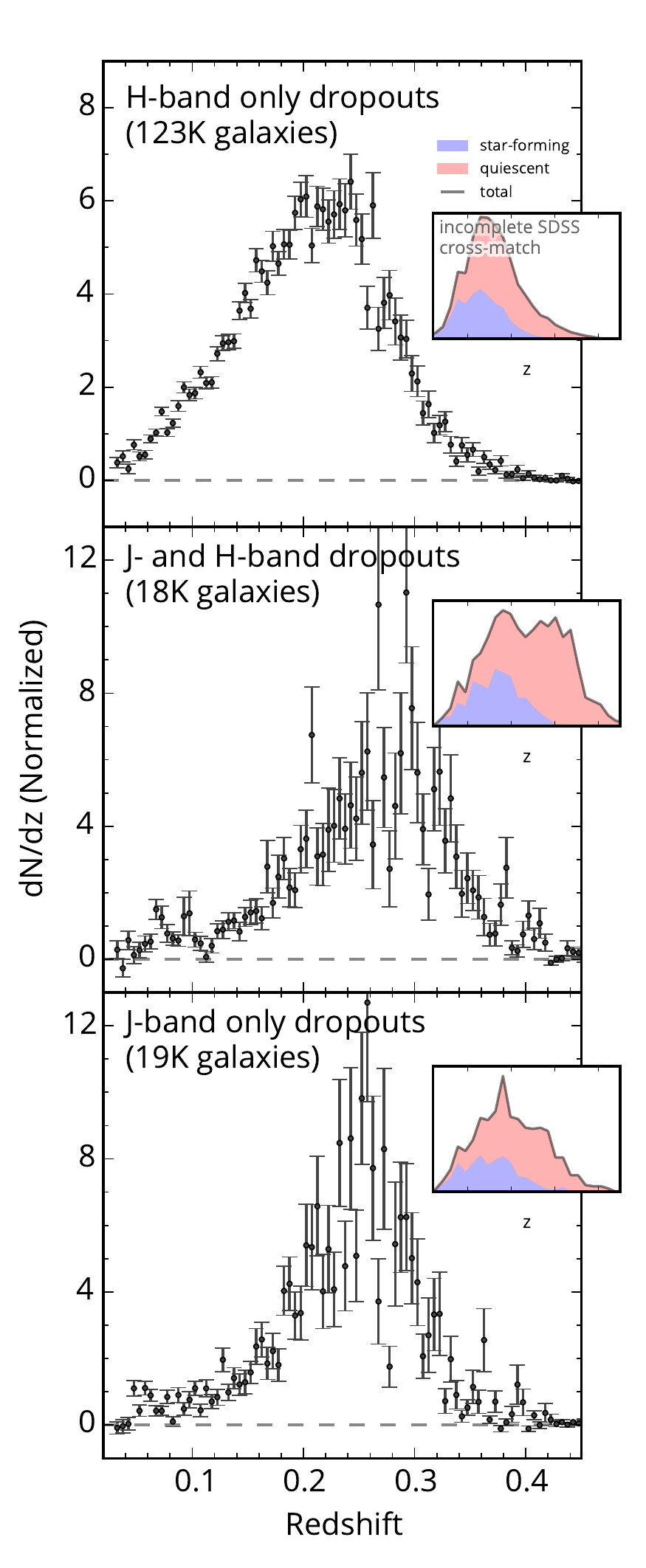}
\caption{
The clustering redshift distributions of H-band dropouts (\emph{top}), J- and H-band dropouts (\emph{middle}) and J-band dropouts (\emph{bottom}). The insets indicate the distributions of SDSS spectroscopic galaxies cross-matched to the 2MASS dropout populations. We separate the samples into star forming (sSFR $> 10^{-11}$ yr$^{-1}$; blue), and quiescent (sSFR $< 10^{-11}$ yr$^{-1}$; red) as measured spectroscopically. The SDSS samples are incomplete. 
\label{fig:dropouts}
}
\end{figure}

The clustering redshift method produces distance information for populations of galaxies with limited or missing photometric information. The 2MASS extended source catalogue contains large numbers of sources with magnitudes below the flux limit in one or more bands ($\sim$ 150 000 at $K_s < 14.5$), which we refer to as ``dropouts''. The 2MASS detection criteria are set by the K-band; consequently, most sources in 2MASS have a measured K-band magnitude but may be missing J- and/or H-band photometry. We separate the dropout populations to determine their redshift distribution: those with detections in J- and K-band (H-band Only dropouts), those with H- and K-band detections (J-band Only dropouts), and those with only K-band detections (J- and H- dropouts). These sources, with only upper limits to their flux in one or more bands, are either challenging to or cannot be characterised by photometric redshift methods. We present the clustering redshift distribution of these populations in Figure \ref{fig:dropouts}. 

The redshift distributions of these populations have an upper redshift bound consistent with the fully detected sources in the 2MASS extended source catalogue; they have little to no redshift signal beyond $z > 0.4$. Since the dropout populations have photometry below the flux limit of one or more bands, the populations tend to be at higher redshifts ($ 0.2 <\overline{z} < 0.3$) as closer sources would be above the limit in all 2MASS bands; nearby objects analogous to these populations would appear as sources with extreme colours. The composition of the dropout populations would differ greatly if the flux limits of the survey were different.

Cross-matching against SDSS sources, we can explore the composition of these 2MASS dropout populations through the fraction of sources with optical spectroscopy. We note that only 50-70\% of the dropouts have spectroscopic observations within the SDSS footprint, and that at $z > 0.15$, the selection is incomplete and biased towards quiescent galaxies by selection. We compare the spectroscopic cross-matched sources with the clustering redshift distributions of the full populations in the insets of Figure \ref{fig:dropouts}. The cross-matched sample have an overall lower redshift distribution than the entire dropout populations. This is expected since the spectroscopic sources are typically flux limited, thus missing the more distant and fainter sources. Consequently, we expect the incompleteness between the full and cross-matched populations to arise at higher redshifts. We separate the cross-matched samples into star-forming and quiescent populations through specific star formation rates, placing the boundary at sSFR = $10^{-11}$ yr$^{-1}$ as measured by \citet{tremonti04}. We demonstrate that all dropout populations come from a combination of star-forming and quiescent populations, indicating that there is no single spectral feature or emission process that causes any given dropout population. 

While the higher redshift population of spectroscopic sources will be biased towards quiescent galaxies, the lower redshift population ($z < 0.1$) will be complete at the magnitudes of the sources in the optical. At low redshift, the sources are dominated by star forming galaxies, with quiescent galaxies contributing a larger fraction as redshift increases. 

Star forming galaxies will have the strongest emission lines, such as the Paschen lines that dominate the NIR spectra, leading to the largest deviation from continuum-dominated colours (Figure \ref{fig:sed}). For fainter sources, this potentially leads to a dropout in one or more of the bands. In particular, the redshift distribution of star-forming galaxies from the spectroscopic cross-match is consistent with the Paschen-$\alpha$ emission line, the strongest in the NIR, being redshifted through the K-band filter. 

For quiescent galaxies, we infer that the differences in flux limits between the J-, H-, and K-bands combined with the galaxy's continuum emission lead to galaxies falling out of one or more bands. This can be caused by the galaxy's extinction and star formation history, all of which imprint an SED-altering signal on the continuum emission of a quiescent galaxy.

\subsection{Global Redshift Distribution}

\begin{figure*}[ht]
\center
\includegraphics[scale=1.0]{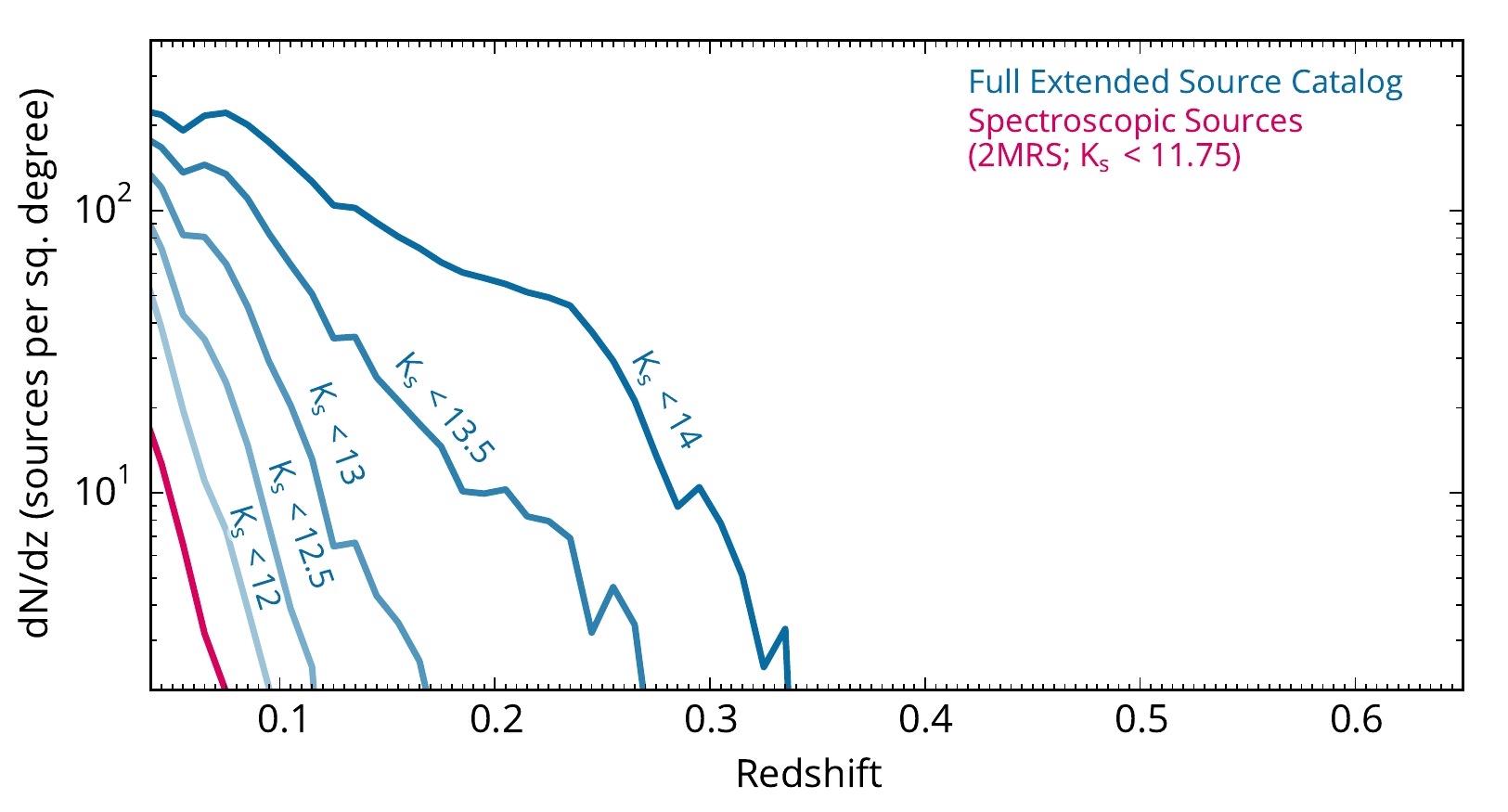}
\caption{
	The total redshift distribution of the 2MASS extended source catalogue as a function of magnitude (in blue). The redshift distribution of 2MRS sources is shown for comparison in the magenta. \label{fig:totdist}
}
\end{figure*}

We now estimate the full redshift distribution of the 2MASS extended source catalogue by summing the distributions of each colour-magnitude selected sample, along with the dropout populations, and weighting the population by the number of sources in each sample. We present this global redshift distribution in Figure~\ref{fig:totdist}. We break the distribution into flux-limited samples to show the magnitude evolution of the catalogue. 
While the 2MASS redshift survey, which roughly corresponds to a magnitude limit of K$_{\textrm s} = 11.75$, probes only $4 \times 10^4$ sources extending to  $z\lesssim0.05$, our clustering redshift analysis allows us to measure redshift distributions for the entire extended source catalogue (over $10^{6}$ objects), which are found to span a redshift range reaching $z\sim0.35$. Our redshift distribution estimates include the contributions of both sources without SDSS spectroscopy, corresponding to about 10\% of the extended source catalogue, as well as near-infrared dropouts which comprise another 10\% of the total sample. Our analysis also allows us to show how the shape of the redshift distribution changes with limiting magnitude, as indicated in Figure~\ref{fig:totdist}.

\section{Clustering Redshifts of\\ 2MASS \emph{point} sources}

We now estimate the clustering redshift distribution for the 2MASS point source catalogue, with $K_S<14$. We use this magnitude limit to remain consistent with the analysis of the extended sources. With a PSF of three arcseconds, we expect a significant fraction of galaxies to be unresolved. The clustering redshift method can be used to (a) determine the fraction of extragalactic sources in the point source catalogue and (b) estimate their redshift distribution.

\subsection{Method}

Galactic stars are not expected to give rise to any spatial cross-correlation signal with extragalactic sources. The set of cross-correlations with references objects can, therefore, be used to probe the redshift distribution of the extragalactic contribution of the sample (galaxies and/or quasars). However, in order to estimate the \emph{number} of extragalactic objects, we can no longer rely on the normalization described in Equation \ref{eq:normalization} since the total number of sources includes both extragalactic and Galactic objects. To estimate the fraction of galaxies $f_{\rm gal}$ in the point source catalogue, we can take several approaches. We list them below, ordered by the amount of external information or assumptions required:
\begin{itemize}
\item The number of sources contributing to the measured overdensity contains information about the number of extragalactic objects. If one considers angular apertures smaller than the mean separation between reference objects, the excess quantity of unknown objects around each reference source provides us with a minimum estimate for the number of extragalactic objects within the sample.
Taking into account the size of the unknown sample, this can then be used to estimate a minimum $f_{\rm gal}$ without the use of any assumption.
If the reference sample covers a wide enough redshift range, this technique can be used to test whether a given sample contains extragalactic sources.
\item Alternatively, assuming that the redshift distribution of the sources is a smooth function, its measured scatter can be used to infer the number of objects contributing to the signal. For example, if $f_{\rm gal}=1$ then all objects in the sample will contribute to the measured cross-correlations as a function of redshift, and consequently, the relative Poisson noise on the estimated $\d N/\d z$ will be low. If, for the same redshift distribution, $f_{\rm gal}$ decreases, then a smaller number of extragalactic objects will contribute to the cross-correlation signal and its scatter will increase systematically. Information on the fraction $f_{\rm gal}$ of sources contributing to the clustering signal can therefore be extracted from the measured scatter of the inferred redshift.
\item 
To get a rough estimate of the fraction of extragalactic objects $f_{\rm gal}$ in the sample, one can compare the amplitude of the measured cross-correlation function to the typical amplitude of galaxy correlation functions of similar sources at a similar redshift. The ratio between the two can be used as an indicator of the dilution factor due to the presence of Galactic sources in the sample which do not contribute to any clustering signal. The dilution factor can be converted into the fraction of sources ($1 - f_{\rm gal}$) that do not contribute to the correlation signal. The accuracy of this approach is limited by the lack of knowledge of the clustering amplitude (or bias) of the extragalactic objects from the unknown sample, which depends on galaxy type and redshift.
\item 
If one can obtain some information on the redshift dependence of the clustering amplitude (or bias) of the unknown sources, one can more precisely relate measured cross-correlation functions to excess number counts. Assuming that the extragalactic sources probed by the point source catalogue with given colours are of similar type as the objects probed by the extended source catalogue, one can first estimate the amplitude of their spatial correlations with the reference sample normalised by the expected biases, which we denote $w'$. If the biases are properly estimated, we expect the ratio $r$ between the number of extragalactic objects contributing to the cross-correlation signal and $w'$ to be roughly constant. One can first verifying that $r$ depends weakly on photometric parameters and redshift. Then, one can use its value to convert the measured correlation functions for the point source catalogue to number counts and therefore obtain an estimate of $f_{\rm gal}$.
\end{itemize}

\subsection{Application to 2MASS}

\begin{figure*}
\center
\includegraphics[scale=1.0]{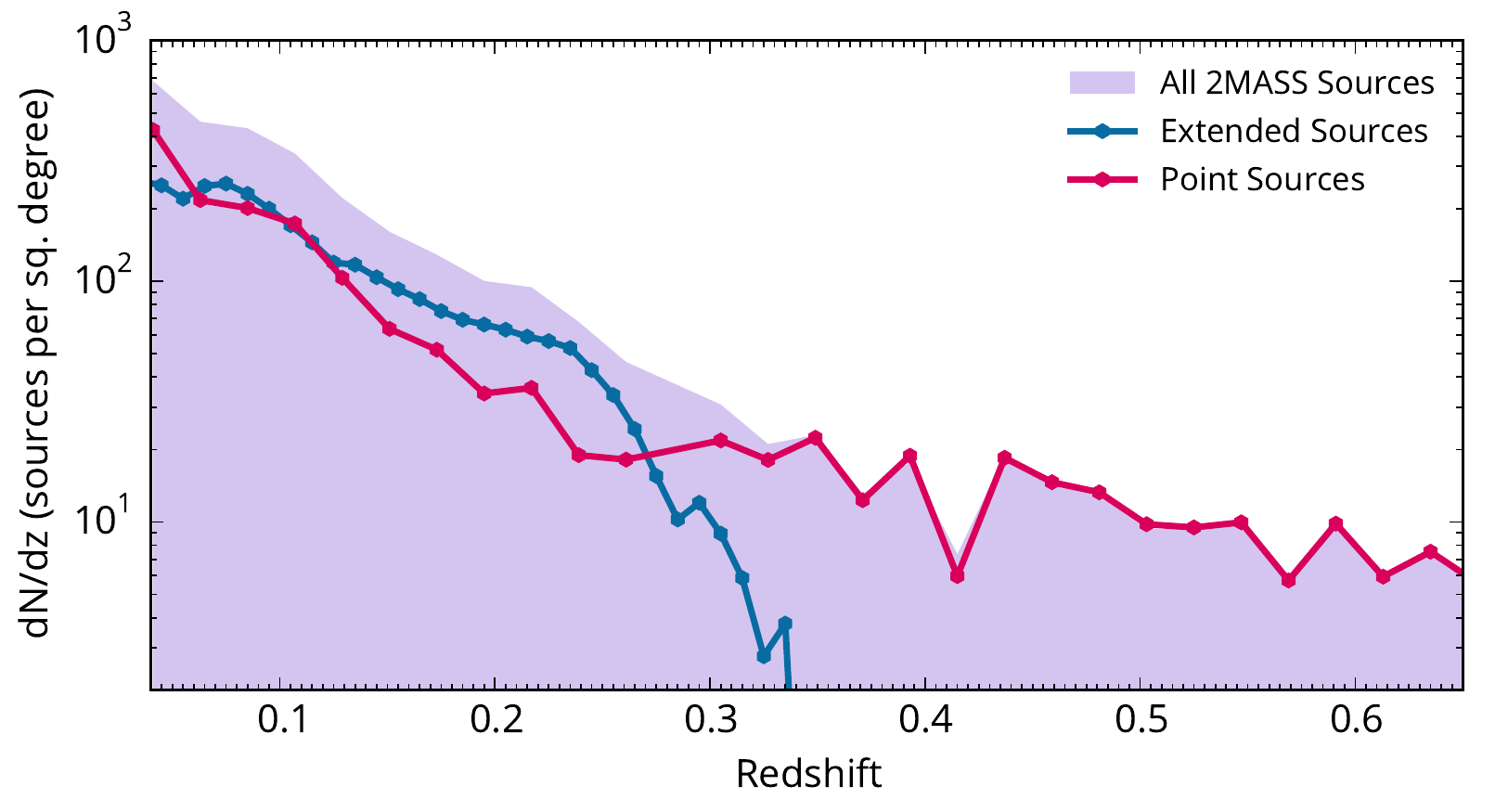}
\caption{
	 The clustering redshift distribution of the 2MASS point sources is presented in magenta with the total redshift distribution of the XSC is shown in blue for comparison. There are as many extragalactic sources in the point source catalogue as there are in the XSC. In purple, we present the total redshift distribution of all extragalactic sources in 2MASS. One noise-dominated point at $z=0.28$ on the point source distribution has been filtered and interpolated over. \label{fig:pscdist}
}
\end{figure*}

Our goal here is only to provide a rough estimate of the fraction $f_{\rm gal}$ of galaxies in the point source catalogue. A precise characterization of $f_{\rm gal}$ and comparison between different estimation methods is beyond the scope of the present analysis. Among the different approaches presented above we will use the latter one; we will assume that the redshift evolution of the clustering properties of the point and extended sources are similar. We note that violations of this assumption only weakly affect the final redshift distribution estimate.

Before applying the corresponding estimator, we select a region of the sky that is more favourable to our task. The fraction of extragalactic sources present in the point source catalogue is expected to be a strong function of galactic latitude. To minimise the stellar contribution, we limit our clustering redshift analysis to a high-galactic latitude footprint defined by:
\begin{eqnarray}
160\degr < &\alpha& < 180\degr\nonumber\\
20\degr < &\delta& <  5\degr
\end{eqnarray}
chosen for computational convienence. This area corresponds to a Galactic latitude of about 65\degr. The noise level of this measurement is expected to be greater than that obtained with the extended source catalogue, as Galactic sources only contribute to shot noise. To maintain a comparable signal-to-noise ratio, we use larger redshift bins for this measurement than in in the previous results ($\Delta z = 0.02$).

For each $(J-H,H-K)$ colour cell of the 2MASS extended source dataset ($i_c$, see \S\ref{sect:colourspace}), we first estimate $w'_{xsc}$,
the clustering amplitude with our reference objects normalised by the expected biases:
\begin{equation}
w'_{xsc}(z, i_c) = \frac{w_{ur}(z, i_c)}{ \overline{b_u}(z)\, \overline{b_r}(z)}\;,
\end{equation}
where $\overline{b_u}$ and $\overline{b_r}$ are the galaxy biases of the unknown and reference samples. As discussed in \S\ref{subsect:clustz}, we use a measured bias for the reference sample, whereas we assume $\d\log \overline{b}/\d \log z=1$ for the 2MASS unknown samples, which we showed to be a reasonable assumption in \S\ref{subsect:2mrs} and \S\ref{sect:colourspace}. 
Next, we compute the ratio between the number counts of unknown sources and the normalised clustering amplitude:
\begin{equation}
r_{\rm xsc}(z,i_c)\equiv 
\frac{N(z, i_c)/N_{tot}(i_c)}{w'_{xsc}(z,i_c)}\;.
\end{equation}
As expected, we can verify that the value of $r(z, i_c)$ is a relatively weak function of colour and redshift: over all the colour cells and redshift bins for which a clustering signal is detectable, we find $\langle r \rangle=1.05$ and $\sigma_r = 0.25$. Using the average value of $r$, we can then convert the measured clustering amplitude as a function of redshift for each colour cell to obtain an estimate of the fraction of galaxies in the point source catalogue. As $r$ appears to be weakly colour dependent, we can directly add the contribution from all redshift bins:
\begin{equation}
f_{gal} =  \langle r \rangle \times \sum_j w_{psc}'(z_j)\; \Delta z_j\;.
\end{equation}

Over the entire redshift range provided by the spectroscopic reference sample (0.03 $<\,z\,<$ 0.8), we estimate that 
\begin{equation}
f_{\rm gal} \simeq 0.10\;.
\end{equation}
This fraction is relatively low. However, given that the point source catalogue density (in this high-galactic latitude region, after removing sources identified as extended) is 337~obj/deg$^{2}$, this leads to an extragalactic point source density of 33~obj/deg$^{2}$. \emph{This source density is as high as that of the extended source catalogue} (31~obj/deg$^{2}$). Unlike galactic sources, we anticipate no variation in the density of the extragalactic sources over the sky. We can compare our clustering-based estimate of $f_{\rm gal}$ to a more direct estimate of this value by computing the fraction of 2MASS point sources classified as extended objects (galaxies) in SDSS. Doing so, we find that 
\begin{equation}
f_{\rm gal} \sim 0.07\;\textrm{ (using SDSS photometry)},
\end{equation}
a number in good agreement with the above estimate. We note that this estimate is based on optical detections and is consequently only a lower limit on the intrinsic value. While our method does not determine which of the galaxies in the catalogue are extragalactic, there are approaches that seek to use information from other surveys to conduct this separation, such as \citet{kovacs15}.

Having estimated $f_{\rm gal}$, we can now characterise the redshift distribution of the extragalactic sources present in the point source catalogue. To do so, we measure the set of spatial cross-correlations between all sources in the point source catalogue with the reference spectroscopic sample introduced in~\ref{sect:colourspace}. 

We present the estimated redshift distribution of both 2MASS extended and point sources with K$_s$ $<$ 14 in Figure \ref{fig:pscdist}. We find a non-zero signal
\footnote{Among the 35 $\d N/\d z$ estimates for the point source catalogue, one redshift bin at $z = 0.28$ led to a negative value, which is unphysical. This point appears to be an outlier in our entire analysis and might be due to a measurement artefact in the photometric dataset, possibly caused inhomogeneities in the on sky source distribution. We decided to exclude it in the final characterization of the redshift distribution.}
from the lowest redshifts up to $z\sim0.8$. As expected, the noise level of the measurement is significantly greater than the earlier work with the extended source catalogue. We find the bulk of the extragalactic point sources to have a redshift distribution similar to that of the extended source catalogue\footnote{The data of the full redshift distributions is available at \url{http://www.pha.jhu.edu/~mubdi/2massz}}. This indicates that, even at low redshift, the extended source catalog alone is incomplete in sampling the total galaxy population. In addition, we observe a tail extending significantly beyond the highest redshifts of the extended sources around $z=0.3$. This maximum redshift is therefore not limited by brightness considerations but that of angular size. Adding the contribution from both the extended and point source catalogues, we find a relatively smooth redshift distribution continuously declining from the lowest redshift probed by our analysis, $z=0.03$ up to $z\sim0.7$.
We note that the cosmic volume probed by the point source catalogue is about ten times larger than that of the extended source catalogue.

\section{Conclusions}
\label{sect:conclusion}

Using clustering-based redshift inference, we have explored the redshift distribution of all sources from the Two-Micron All-Sky Survey survey, through the extended and point source catalogue containing 1.6 and 470 million objects across the entire sky. As its near-infrared photometry primarily probes the featureless Rayleigh-Jeans tail of galaxy spectral energy distributions, colour-based redshift estimation is uninformative.

We applied the clustering redshift technique following the implementation presented in \citet{rahman15} to about 5,000 square degrees of 2MASS data. Our main results are as follows:
\begin{itemize}
\item We first demonstrated the robustness of our clustering redshift technique by reproducing the redshift distribution of sources with $K_s < 11.75$, for which the 2MASS Redshift Survey provides complete spectroscopic redshift measurements.
\item We measured the redshift distributions of 2MASS extended sources down to $K_s=14.0$ as a function of their $J-H$ and $H-K_s$ colours. A comparison to spectroscopic redshifts from SDSS optical data shows excellent agreement. In addition, we presented redshift distribution estimates for near-infrared dropouts in $J$, $H$ and $J\,\&\,H$, for which only limited spectroscopic information is available and colour-based techniques tend to fail. Finally we combine all these subsamples to present an estimate of the global redshift distribution of the 2MASS extended catalogue which displays sources up to $z\sim0.3$.
\item We explored the content of the Point Source Catalogue and, based on clustering information, showed that at high Galactic latitude about 10\% of the sources are extragalactic down to $K_s=14.0$. This implies that the point source catalogue contains about 1.6 million extragalactic objects, i.e., \emph{as many} as in the extended source catalogue. We presented the redshift distribution of these sources, which show galaxies extending to $z\sim0.7$. The Point Source Catalogue therefore provides a full-sky sample of extragalactic sources 
with a cosmic volume about ten times larger than that of the extended source catalogue. It is therefore of potential interest for cosmological experiments such as measurements of the Integrated Sachs-Wolfe effect. %
\item Finally, adding the contributions from both the extended and point source catalogues, we presented the full redshift distribution of the 2MASS survey which showed a relatively smooth distribution continuously declining from the lowest redshift probed by our analysis, $z=0.03$, up to $z\sim0.7$.
\end{itemize}
In addition to characterising the 2MASS survey, this analysis illustrates the potential of clustering redshift estimation. Previous redshift estimates of 2MASS sources have relied on optical observations, which suffer from a selection bias present when near-infrared sources are not detected in the optical. In contrast, clustering-redshift estimation can be performed regardless of the optical properties of the sources and can directly be used to infer the redshift distribution of the entire 2MASS dataset.
When possible, we have presented direct comparisons to spectroscopic redshift distributions which show excellent agreement, hence validating our method and its implementation. Our redshift distribution estimates as a function of near-infrared colours, valid for the entire sky, are made available through a webpage. This work further demonstrates the power and potential of the clustering redshift technique.

\section*{acknowledgments}

The authors thank A. Mendez for useful discussions. The authors also thank S. Schmidt, T. Lan, and M. Bilicki for extensive comments on the manuscript. This work is supported by NASA grant 12-ADAP12-0270 and National Science Foundation grant AST-1313302. This publication makes use of data products from the Two Micron All Sky Survey, which is a joint project of the University of Massachusetts and the Infrared Processing and Analysis Center/California Institute of Technology, funded by the National Aeronautics and Space Administration and the National Science Foundation.

Funding for SDSS-III has been provided by the Alfred P. Sloan Foundation, the Participating Institutions, the National Science Foundation, and the U.S. Department of Energy Office of Science. The SDSS-III web site is http://www.sdss3.org/.

SDSS-III is managed by the Astrophysical Research Consortium for the Participating Institutions of the SDSS-III Collaboration including the University of Arizona, the Brazilian Participation Group, Brookhaven National Laboratory, University of Cambridge, Carnegie Mellon University, University of Florida, the French Participation Group, the German Participation Group, Harvard University, the Instituto de Astrofisica de Canarias, the Michigan State/Notre Dame/JINA Participation Group, Johns Hopkins University, Lawrence Berkeley National Laboratory, Max Planck Institute for Astrophysics, Max Planck Institute for Extraterrestrial Physics, New Mexico State University, New York University, Ohio State University, Pennsylvania State University, University of Portsmouth, Princeton University, the Spanish Participation Group, University of Tokyo, University of Utah, Vanderbilt University, University of Virginia, University of Washington, and Yale University. 

Facilities: \facility{Sloan}, \facility{2MASS}

\newpage
\bibliography{rs-2mass}

\begin{thebibliography}{29}
\expandafter\ifx\csname natexlab\endcsname\relax\def\natexlab#1{#1}\fi

\bibitem[{{Bilicki} {et~al.}(2014){Bilicki}, {Jarrett}, {Peacock}, {Cluver}, \&
  {Steward}}]{bilicki14}
{Bilicki}, M., {Jarrett}, T.~H., {Peacock}, J.~A., {Cluver}, M.~E., \&
  {Steward}, L. 2014, \apjs, 210, 9, 9

\bibitem[{{Eisenstein} {et~al.}(2001){Eisenstein}, {Annis}, {Gunn}, {Szalay},
  {Connolly}, {Nichol}, {Bahcall}, {Bernardi}, {Burles}, {Castander},
  {Fukugita}, {Hogg}, {Ivezi{\'c}}, {Knapp}, {Lupton}, {Narayanan}, {Postman},
  {Reichart}, {Richmond}, {Schneider}, {Schlegel}, {Strauss}, {SubbaRao},
  {Tucker}, {Vanden Berk}, {Vogeley}, {Weinberg}, \& {Yanny}}]{eisenstein01}
{Eisenstein}, D.~J., {Annis}, J., {Gunn}, J.~E., {et~al.} 2001, \aj, 122, 2267,
  2267

\bibitem[{{Francis} \& {Peacock}(2010)}]{francis2010}
{Francis}, C.~L., \& {Peacock}, J.~A. 2010, \mnras, 406, 2, 2

\bibitem[{{Hambly} {et~al.}(2001){Hambly}, {MacGillivray}, {Read}, {Tritton},
  {Thomson}, {Kelly}, {Morgan}, {Smith}, {Driver}, {Williamson}, {Parker},
  {Hawkins}, {Williams}, \& {Lawrence}}]{hambly01}
{Hambly}, N.~C., {MacGillivray}, H.~T., {Read}, M.~A., {et~al.} 2001, \mnras,
  326, 1279, 1279

\bibitem[{{Ho} {et~al.}(2008){Ho}, {Hirata}, {Padmanabhan}, {Seljak}, \&
  {Bahcall}}]{ho08}
{Ho}, S., {Hirata}, C., {Padmanabhan}, N., {Seljak}, U., \& {Bahcall}, N. 2008,
  \prd, 78, 043519, 043519

\bibitem[{{Huchra} {et~al.}(2012){Huchra}, {Macri}, {Masters}, {Jarrett},
  {Berlind}, {Calkins}, {Crook}, {Cutri}, {Erdo{\v g}du}, {Falco}, {George},
  {Hutcheson}, {Lahav}, {Mader}, {Mink}, {Martimbeau}, {Schneider},
  {Skrutskie}, {Tokarz}, \& {Westover}}]{huchra12}
{Huchra}, J.~P., {Macri}, L.~M., {Masters}, K.~L., {et~al.} 2012, \apjs, 199,
  26, 26

\bibitem[{{Jarrett}(2004)}]{jarrett04}
{Jarrett}, T. 2004, PASA, 21, 396, 396

\bibitem[{{Jarrett} {et~al.}(2000){Jarrett}, {Chester}, {Cutri}, {Schneider},
  {Skrutskie}, \& {Huchra}}]{jarrett00}
{Jarrett}, T.~H., {Chester}, T., {Cutri}, R., {et~al.} 2000, \aj, 119, 2498,
  2498

\bibitem[{{Jones} {et~al.}(2009){Jones}, {Read}, {Saunders}, {Colless},
  {Jarrett}, {Parker}, {Fairall}, {Mauch}, {Sadler}, {Watson}, {Burton},
  {Campbell}, {Cass}, {Croom}, {Dawe}, {Fiegert}, {Frankcombe}, {Hartley},
  {Huchra}, {James}, {Kirby}, {Lahav}, {Lucey}, {Mamon}, {Moore}, {Peterson},
  {Prior}, {Proust}, {Russell}, {Safouris}, {Wakamatsu}, {Westra}, \&
  {Williams}}]{jones09}
{Jones}, D.~H., {Read}, M.~A., {Saunders}, W., {et~al.} 2009, \mnras, 399, 683,
  683

\bibitem[{{Kochanek} {et~al.}(2003){Kochanek}, {White}, {Huchra}, {Macri},
  {Jarrett}, {Schneider}, \& {Mader}}]{kochanek03}
{Kochanek}, C.~S., {White}, M., {Huchra}, J., {et~al.} 2003, \apj, 585, 161,
  161

\bibitem[{{Kotulla} {et~al.}(2009){Kotulla}, {Fritze}, {Weilbacher}, \&
  {Anders}}]{kotulla09}
{Kotulla}, R., {Fritze}, U., {Weilbacher}, P., \& {Anders}, P. 2009, \mnras,
  396, 462, 462

\bibitem[{{Kov{\'a}cs} \& {Szapudi}(2015)}]{kovacs15}
{Kov{\'a}cs}, A., \& {Szapudi}, I. 2015, \mnras, 448, 1305, 1305

\bibitem[{{Landy} {et~al.}(1996){Landy}, {Szalay}, \& {Koo}}]{landy96}
{Landy}, S.~D., {Szalay}, A.~S., \& {Koo}, D.~C. 1996, \apj, 460, 94, 94

\bibitem[{{McQuinn} \& {White}(2013)}]{mcquinn13}
{McQuinn}, M., \& {White}, M. 2013, ArXiv e-prints, arXiv:1302.0857

\bibitem[{{M{\'e}nard} {et~al.}(2013){M{\'e}nard}, {Scranton}, {Schmidt},
  {Morrison}, {Jeong}, {Budavari}, \& {Rahman}}]{menard13}
{M{\'e}nard}, B., {Scranton}, R., {Schmidt}, S., {et~al.} 2013, ArXiv e-prints,
  arXiv:1303.4722

\bibitem[{{Newman}(2008)}]{newman08}
{Newman}, J.~A. 2008, \apj, 684, 88, 88

\bibitem[{{Padmanabhan} {et~al.}(2012){Padmanabhan}, {Xu}, {Eisenstein},
  {Scalzo}, {Cuesta}, {Mehta}, \& {Kazin}}]{padman12}
{Padmanabhan}, N., {Xu}, X., {Eisenstein}, D.~J., {et~al.} 2012, \mnras, 427,
  2132, 2132

\bibitem[{{Rahman} {et~al.}(2015{\natexlab{a}}){Rahman}, {M{\'e}nard},
  {Scranton}, {Schmidt}, \& {Morrison}}]{rahman15}
{Rahman}, M., {M{\'e}nard}, B., {Scranton}, R., {Schmidt}, S.~J., \&
  {Morrison}, C.~B. 2015{\natexlab{a}}, \mnras, 447, 3500, 3500

\bibitem[{{Rahman} {et~al.}(2015{\natexlab{b}}){Rahman}, {Mendez},
  {M{\'e}nard}, {Scranton}, {Schmidt}, {Morrison}, \&
  {Budav{\'a}ri}}]{rahman15b}
{Rahman}, M., {Mendez}, A.~J., {M{\'e}nard}, B., {et~al.} 2015{\natexlab{b}},
  ArXiv e-prints, arXiv:1512.03057

\bibitem[{{Schmidt} {et~al.}(2013){Schmidt}, {M{\'e}nard}, {Scranton},
  {Morrison}, \& {McBride}}]{schmidt13}
{Schmidt}, S.~J., {M{\'e}nard}, B., {Scranton}, R., {Morrison}, C., \&
  {McBride}, C.~K. 2013, \mnras, arXiv:1303.0292

\bibitem[{{Seldner} \& {Peebles}(1979)}]{seldner79}
{Seldner}, M., \& {Peebles}, P.~J.~E. 1979, \apj, 227, 30, 30

\bibitem[{{Skrutskie} {et~al.}(2006){Skrutskie}, {Cutri}, {Stiening},
  {Weinberg}, {Schneider}, {Carpenter}, {Beichman}, {Capps}, {Chester},
  {Elias}, {Huchra}, {Liebert}, {Lonsdale}, {Monet}, {Price}, {Seitzer},
  {Jarrett}, {Kirkpatrick}, {Gizis}, {Howard}, {Evans}, {Fowler}, {Fullmer},
  {Hurt}, {Light}, {Kopan}, {Marsh}, {McCallon}, {Tam}, {Van Dyk}, \&
  {Wheelock}}]{skrutskie06}
{Skrutskie}, M.~F., {Cutri}, R.~M., {Stiening}, R., {et~al.} 2006, \aj, 131,
  1163, 1163

\bibitem[{{Strauss} {et~al.}(2002){Strauss}, {Weinberg}, {Lupton}, {Narayanan},
  {Annis}, {Bernardi}, {Blanton}, {Burles}, {Connolly}, {Dalcanton}, {Doi},
  {Eisenstein}, {Frieman}, {Fukugita}, {Gunn}, {Ivezi{\'c}}, {Kent}, {Kim},
  {Knapp}, {Kron}, {Munn}, {Newberg}, {Nichol}, {Okamura}, {Quinn}, {Richmond},
  {Schlegel}, {Shimasaku}, {SubbaRao}, {Szalay}, {Vanden Berk}, {Vogeley},
  {Yanny}, {Yasuda}, {York}, \& {Zehavi}}]{strauss02}
{Strauss}, M.~A., {Weinberg}, D.~H., {Lupton}, R.~H., {et~al.} 2002, \aj, 124,
  1810, 1810

\bibitem[{{Tremonti} {et~al.}(2004){Tremonti}, {Heckman}, {Kauffmann},
  {Brinchmann}, {Charlot}, {White}, {Seibert}, {Peng}, {Schlegel}, {Uomoto},
  {Fukugita}, \& {Brinkmann}}]{tremonti04}
{Tremonti}, C.~A., {Heckman}, T.~M., {Kauffmann}, G., {et~al.} 2004, \apj, 613,
  898, 898

\bibitem[{{Wang} {et~al.}(2008){Wang}, {Zhang}, {Liu}, \& {Zhao}}]{wang08}
{Wang}, D., {Zhang}, Y.-X., {Liu}, C., \& {Zhao}, Y.-H. 2008, \cjaa, 8, 119,
  119

\bibitem[{{Wang} {et~al.}(2009){Wang}, {Huang}, \& {Gu}}]{wang09}
{Wang}, T., {Huang}, J.-S., \& {Gu}, Q.-S. 2009, Research in Astronomy and
  Astrophysics, 9, 390, 390

\bibitem[{{Way} {et~al.}(2009){Way}, {Foster}, {Gazis}, \&
  {Srivastava}}]{way09}
{Way}, M.~J., {Foster}, L.~V., {Gazis}, P.~R., \& {Srivastava}, A.~N. 2009,
  \apj, 706, 623, 623

\bibitem[{{Way} \& {Srivastava}(2006)}]{way06}
{Way}, M.~J., \& {Srivastava}, A.~N. 2006, \apj, 647, 102, 102

\bibitem[{{York} {et~al.}(2000){York}, {Adelman}, {Anderson}, {Anderson},
  {Annis}, {Bahcall}, {Bakken}, {Barkhouser}, {Bastian}, {Berman}, {Boroski},
  {Bracker}, {Briegel}, {Briggs}, {Brinkmann}, {Brunner}, {Burles}, {Carey},
  {Carr}, {Castander}, {Chen}, {Colestock}, {Connolly}, {Crocker}, {Csabai},
  {Czarapata}, {Davis}, {Doi}, {Dombeck}, {Eisenstein}, {Ellman}, {Elms},
  {Evans}, {Fan}, {Federwitz}, {Fiscelli}, {Friedman}, {Frieman}, {Fukugita},
  {Gillespie}, {Gunn}, {Gurbani}, {de Haas}, {Haldeman}, {Harris}, {Hayes},
  {Heckman}, {Hennessy}, {Hindsley}, {Holm}, {Holmgren}, {Huang}, {Hull},
  {Husby}, {Ichikawa}, {Ichikawa}, {Ivezi{\'c}}, {Kent}, {Kim}, {Kinney},
  {Klaene}, {Kleinman}, {Kleinman}, {Knapp}, {Korienek}, {Kron}, {Kunszt},
  {Lamb}, {Lee}, {Leger}, {Limmongkol}, {Lindenmeyer}, {Long}, {Loomis},
  {Loveday}, {Lucinio}, {Lupton}, {MacKinnon}, {Mannery}, {Mantsch}, {Margon},
  {McGehee}, {McKay}, {Meiksin}, {Merelli}, {Monet}, {Munn}, {Narayanan},
  {Nash}, {Neilsen}, {Neswold}, {Newberg}, {Nichol}, {Nicinski}, {Nonino},
  {Okada}, {Okamura}, {Ostriker}, {Owen}, {Pauls}, {Peoples}, {Peterson},
  {Petravick}, {Pier}, {Pope}, {Pordes}, {Prosapio}, {Rechenmacher}, {Quinn},
  {Richards}, {Richmond}, {Rivetta}, {Rockosi}, {Ruthmansdorfer}, {Sandford},
  {Schlegel}, {Schneider}, {Sekiguchi}, {Sergey}, {Shimasaku}, {Siegmund},
  {Smee}, {Smith}, {Snedden}, {Stone}, {Stoughton}, {Strauss}, {Stubbs},
  {SubbaRao}, {Szalay}, {Szapudi}, {Szokoly}, {Thakar}, {Tremonti}, {Tucker},
  {Uomoto}, {Vanden Berk}, {Vogeley}, {Waddell}, {Wang}, {Watanabe},
  {Weinberg}, {Yanny}, {Yasuda}, \& {SDSS Collaboration}}]{york00}
{York}, D.~G., {Adelman}, J., {Anderson}, Jr., J.~E., {et~al.} 2000, \aj, 120,
  1579, 1579

\end{thebibliography}

\end{document}